# New Design Study and Related Experimental Program for the LCLS RF Photoinjector†

M. Ferrario[a], P. R. Bolton[b], J. E. Clendenin[b], D. H. Dowell[c], S. M. Gierman[b], M. E. Hernandez[b], D. Nguyen[d], D. T. Palmer[b], J. B. Rosenzweig[e], J. F. Schmerge[b], L. Serafini[f]

[a]*INFN–Laboratori Nazionali di Frascati, Via E. Fermi 40, 00044 Frascati (Roma), Italy*
[b]*Stanford Linear Accelerator Center, Stanford, CA 94309, USA*
[c]*Boeing Physical Sciences Research Center, Seattle, WA 98124, USA*
[d]*Los Alamos National Laboratory , Los Alamos, NM 87545, USA*
[e]*UCLA, Dept. of Physics and Astronomy, Los Angeles, CA 90095, USA*
[f]*INFN–Sezione di Milano, Via Celoria 16, 20133 Milano, Italy*

## Abstract

We report the results of a recent beam dynamics study, motivated by the need to redesign the LCLS photoinjector, that lead to the discovery of a new effective working point for a split RF photoinjector. We consider the emittance compensation regime of a space charge beam: by increasing the solenoid strength, the emittance evolution shows a double minimum behavior in the drifting region. If the booster is located where the relative emittance maximum and the envelope waist occur, the second emittance minimum can be shifted to the booster exit and frozen at a very low level (0.3 mm-mrad for a 1 nC flat top bunch), to the extent that the invariant envelope matching conditions are satisfied. Standing Wave Structures or alternatively Traveling Wave Structures embedded in a Long Solenoid are both candidates as booster linac. A careful measurement of the emittance evolution as a function of position in the drifting region is necessary to verify the computation and to determine experimentally the proper position of the booster cavities. The new design study and supporting experimental program under way at the SLAC Gun Test Facility are discussed.

*Contributed to the*
*7th European Particle Accelerator Conference*
*Vienna, Austria*
*26-30 June 2000*

†Work supported by Department of Energy contract DE-AC03-76SF00515.

# NEW DESIGN STUDY AND RELATED EXPERIMENTAL PROGRAM FOR THE LCLS RF PHOTOINJECTOR

M. Ferrario (INFN-LNF), P. R. Bolton (SLAC),J. E. Clendenin (SLAC),
D. H. Dowell (BOEING), S. M. Gierman (SLAC), M. E. Hernandez (SLAC),
D. Nguyen (LANL), D. T. Palmer (SLAC), J. B. Rosenzweig (UCLA),
J. F. Schmerge (SLAC), L. Serafini (INFN-MI)

*Abstract*

We report the results of a recent beam dynamics study, motivated by the need to redesign the LCLS photoinjector, that lead to the discovery of a new effective working point for a split RF photoinjector. We consider the emittance compensation regime of a space charge beam: by increasing the solenoid strength, the emittance evolution shows a double minimum behavior in the drifting region. If the booster is located where the relative emittance maximum and the envelope waist occur, the second emittance minimum can be shifted to the booster exit and frozen at a very low level (0.3 mm-mrad for a 1 nC flat top bunch), to the extent that the invariant envelope matching conditions are satisfied. Standing Wave Structures or alternatively Traveling Wave Structures embedded in a Long Solenoid are both candidates as booster linac. A careful measurement of the emittance evolution as a function of position in the drifting region is necessary to verify the computation and to determine experimentally the proper position of the booster cavities. The new design study and supporting experimental program under way at the SLAC Gun Test Facility are discussed.

## 1 INTRODUCTION

The proposed Linac Coherent Light Source (LCLS) is an X-ray Free Electron Laser (FEL) that will use the final 15 GeV of the SLAC 3-km linac for the drive beam. The performance of the LCLS in the 1.5 Angstrom regime is predicated on the availability of a 1-nC, 100-A beam at the 160-MeV point with normalised rms transverse emittance of ~1 mm-mrad. An experimental program is underway at the Gun Test Facility (GTF) at SLAC to demonstrate a high-brightness beam meeting the LCLS requirements using an rf photoinjector [1]. The GTF experiment uses a 1.6-cell S-band rf gun developed jointly with BNL and UCLA [2] surrounded by a solenoid just after the gun exit. After a short drift it is followed by a standard SLAC 3-m accelerating section. A transverse normalised rms emittance of 2.4 mm-mrad for a 0.9 nC pulse with 10-ps FWHM Gaussian pulse length has been measured at BNL using a similar configuration [3].

Earlier simulation studies using the multi-particle code PARMELA predicted a transverse normalised emittance of ~1 mm-mrad - thermal emittance not included - for the LCLS photoinjector if a uniform (or even a truncated Gaussian) temporal charge distribution is used [4]. While this result technically meets the LCLS requirements, it leaves no headroom for errors or practical difficulties. Consequently simulation studies have continued with the goal of finding a photoinjector design for the LCLS that would predict a transverse emittance of no more than 0.8 mm-mrad with the thermal emittance included.

## 2 A NEW WORKING POINT FOR A SPLIT PHOTO-INJECTOR

A detailed systematic PARMELA analysis of the beam parameters for a GTF-like system can be found in [5]. Good emittance performances and high peak current at the exit of the device can be obtained with high peak field on the cathode (140 MV/m), extraction phase 35 deg and moderate solenoid field strength (0.3 T) for a 1 nC uniform charge distribution and 10 ps long bunch with 1 mm radius. In the following analysis we will take these parameters as a starting point.

We adopted the semi-analytical code HOMDYN [6] to extend the investigation to the booster matching condition, taking advantage of the fast running capability of the code to explore a wider range of parameters. The invariant envelope [7] matching condition indeed requires the beam to be injected at a laminar waist, $\sigma' = 0$, in a matched accelerating gradient of the TW booster given by:

$$\gamma' = \frac{2}{\sigma_w}\sqrt{\frac{\hat{I}}{2I_0\gamma}}$$

where $I_o$ = 17 kA is the Alfven current. While investigating the envelope and emittance behaviour by scanning the solenoid field strength we noticed an interesting feature. By increasing the solenoid strength the emittance evolution shows a double minimum behaviour in the drifting region. For a given value of the solenoid strength (0.31 T) the envelope waist occurs where the emittance has its relative maximum (z≈1.5 m) as shown by the bold red lines in Figures 1 and 2.

The performance using the new working points [8] relies on this feature of the emittance oscillation. The guess is that if we locate the booster entrance at z ≈ 1.5 and we satisfy the matching condition, the second emittance minimum could be shifted to higher energy and frozen at a lower level, taking advantage of the additional emittance compensation occurring in the booster.

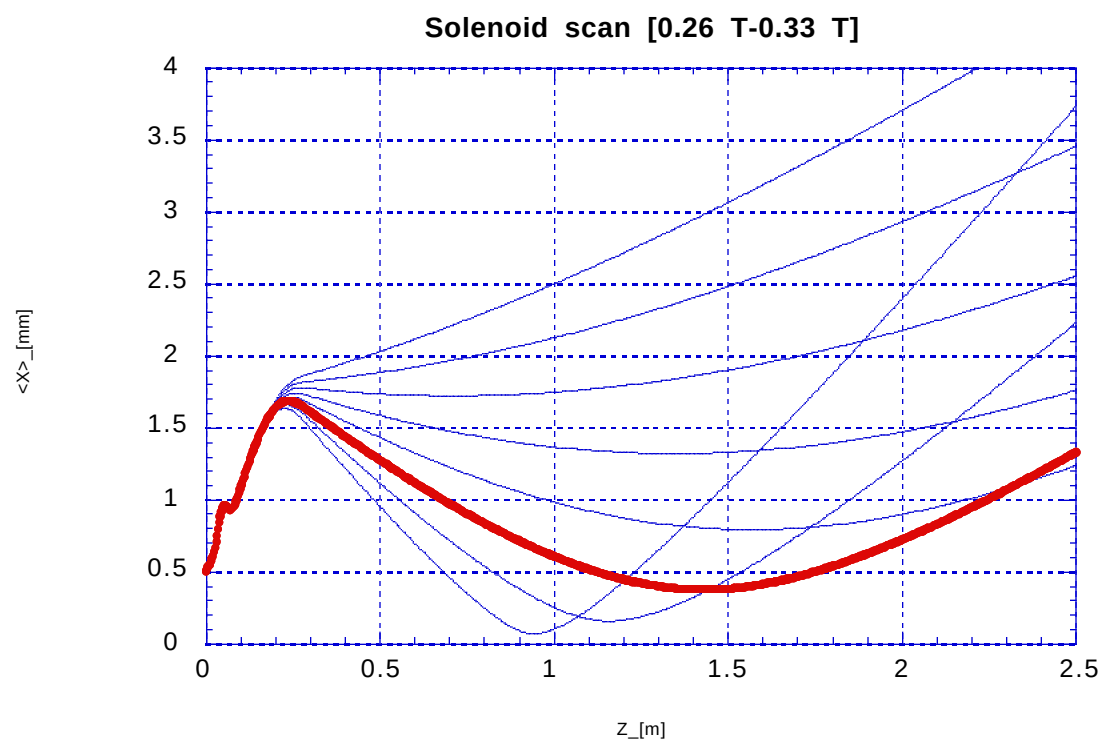

Figure 1: Beam envelope versus z for different solenoid strengths.

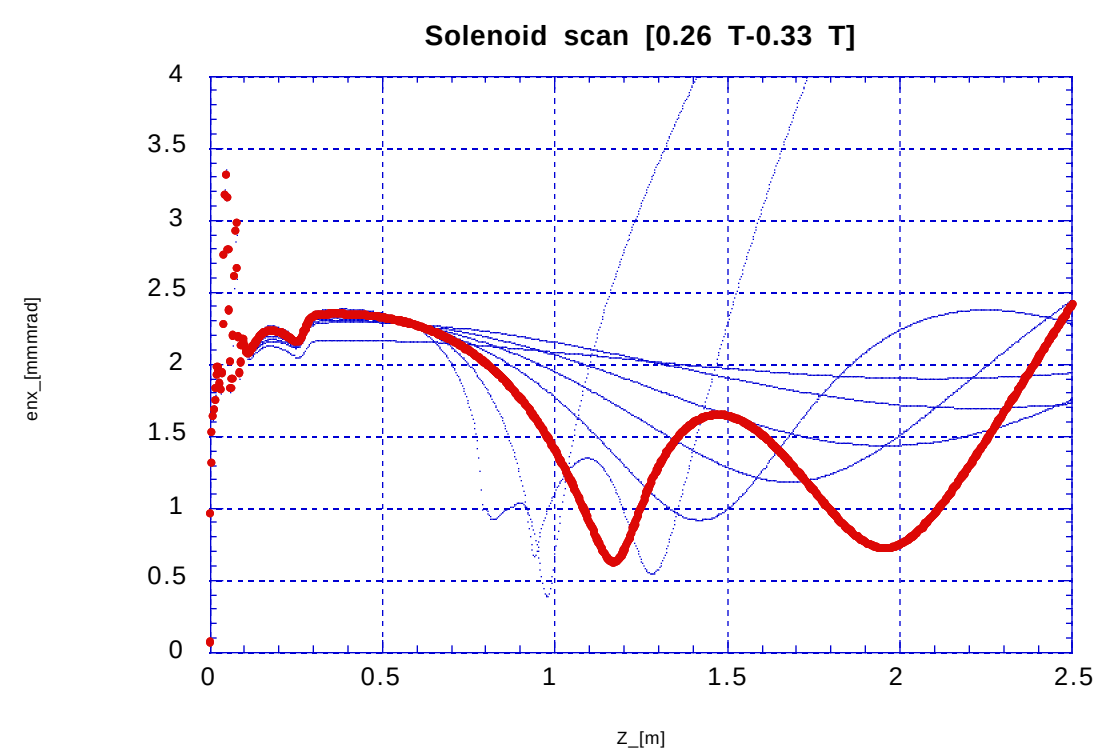

Figure 2: Beam emittance versus z for different solenoid strengths.

## 3 THE LCLS CASE

As a first natural choice for the booster linac we consider a 3 m long constant gradient S-band TW structure, in use at SLAC since the 1960s. To satisfy the first matching condition the booster should be located at z=1.5 m, where the beam laminar waist occurs. Since the rms spot size is $\sigma_x$=0.41 mm, the current averaged over the slices is 96 A, and the average slice energy is 6.4 MeV, the matched accelerating gradient of the TW booster is required to be 35 MV/m. To drive the beam out of the space charge dominated regime we need two SLAC structures resulting in an energy of 216 MeV in an 8 m long injector line, allowing a 0.5 m long drift in between the two structures.

As expected the second emittance minimum, 0.5 mm-mrad, now occurs downstream of the booster structures, see Fig. 3, at z=10 m (we will take this location as a reference position to quote emittance at the injector exit). An injection into the booster 12 degrees off crest is enough to compensate for the energy spread, resulting in a residual rms energy spread at the exit of the booster of 0.2 % as required by the LCLS specifications, see Figure 4. The peak current of 96 A is slightly below the desired value.

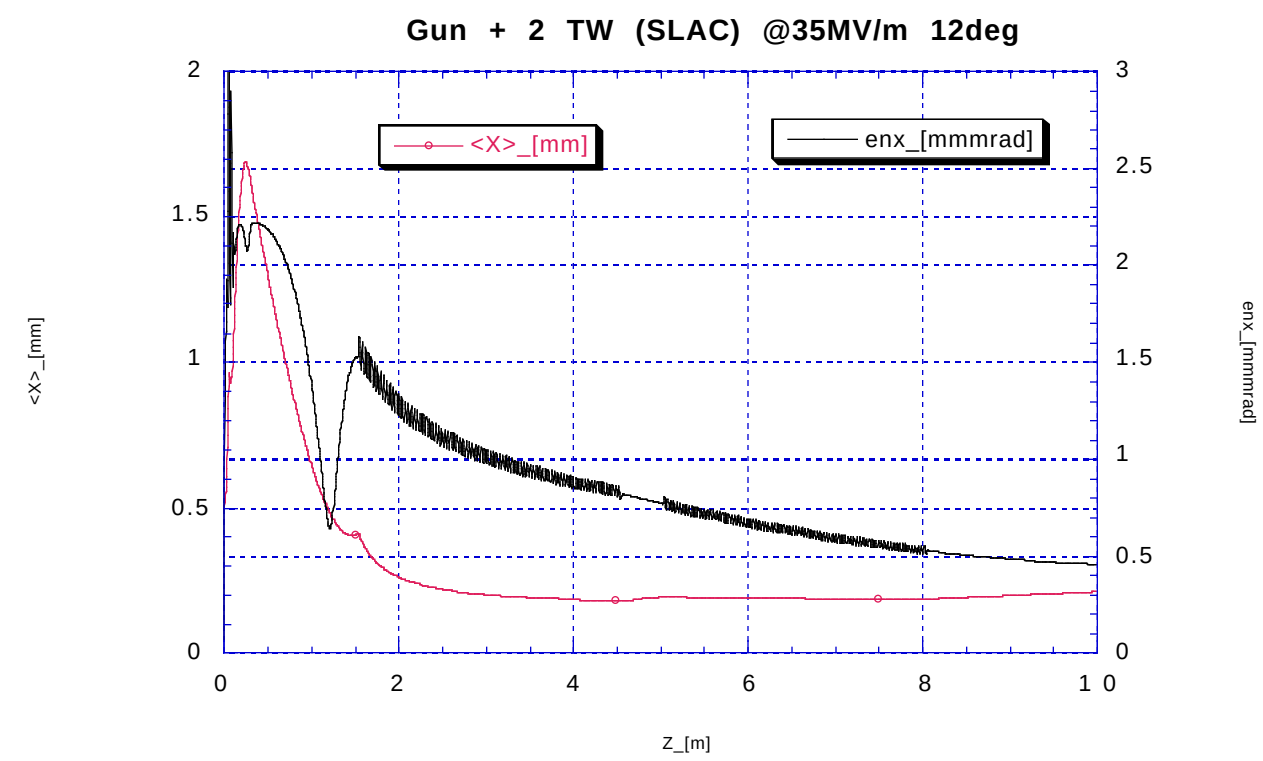

Figure 3: Beam envelope and emittance along the injector beam line.

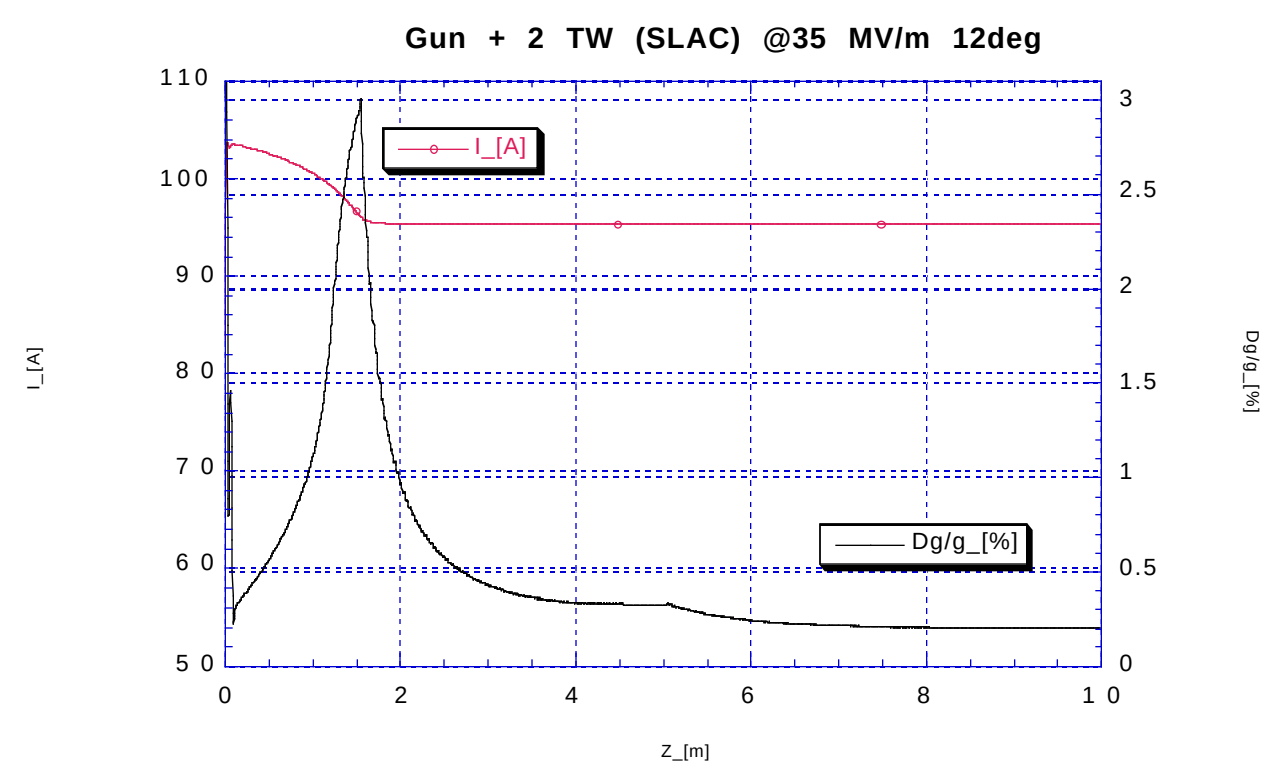

Figure 4: Peak current and rms energy spread along the injector beam line.

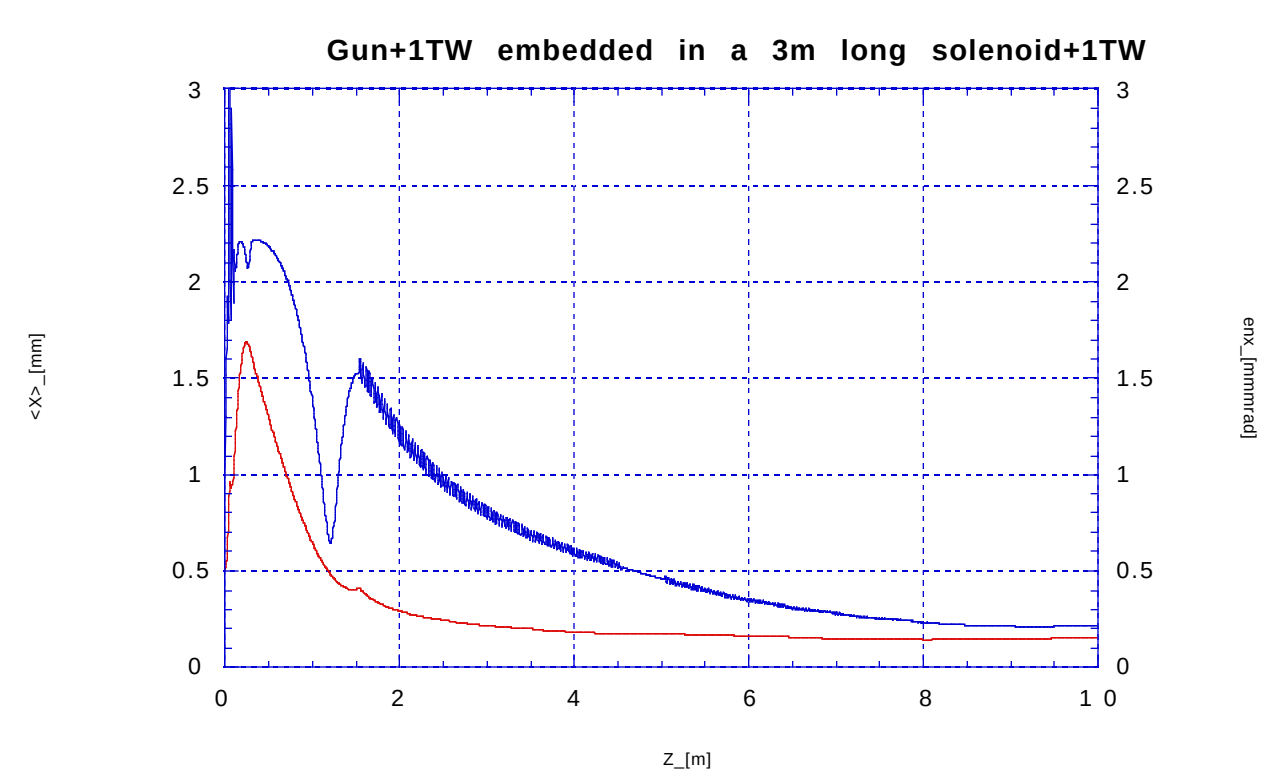

Figure 5: Beam envelope and emittance along the injector beam line.

Despite the good emittance resulting from this design, the necessary gradient to match the beam to the booster exceeds the limit of reliable performance by the SLAC structure.

We then started to look for a lower gradient solution by increasing the focusing properties of the device. This can be done by means of a long solenoid around the first TW structure. Setting the desired accelerating field of the TW section to 25 MV/m, a scanning of the solenoid strength showed a very good working point with B=800 G. Beam dynamics simulation showed a very low emittance value, 0.2 mm mrad, see Fig. 5, fulfilling at the same time the other LCLS requirements.

# 4 PULSE RISE TIME AND THERMAL EMITTANCE EFFECTS

We investigated the emittance degradation due to a realistic pulse rise time by means of PARMELA [9] simulations. The results, seen in Fig. 6, show a very good performance of the new working point up to a 1 psec pulse rise time, the design value.

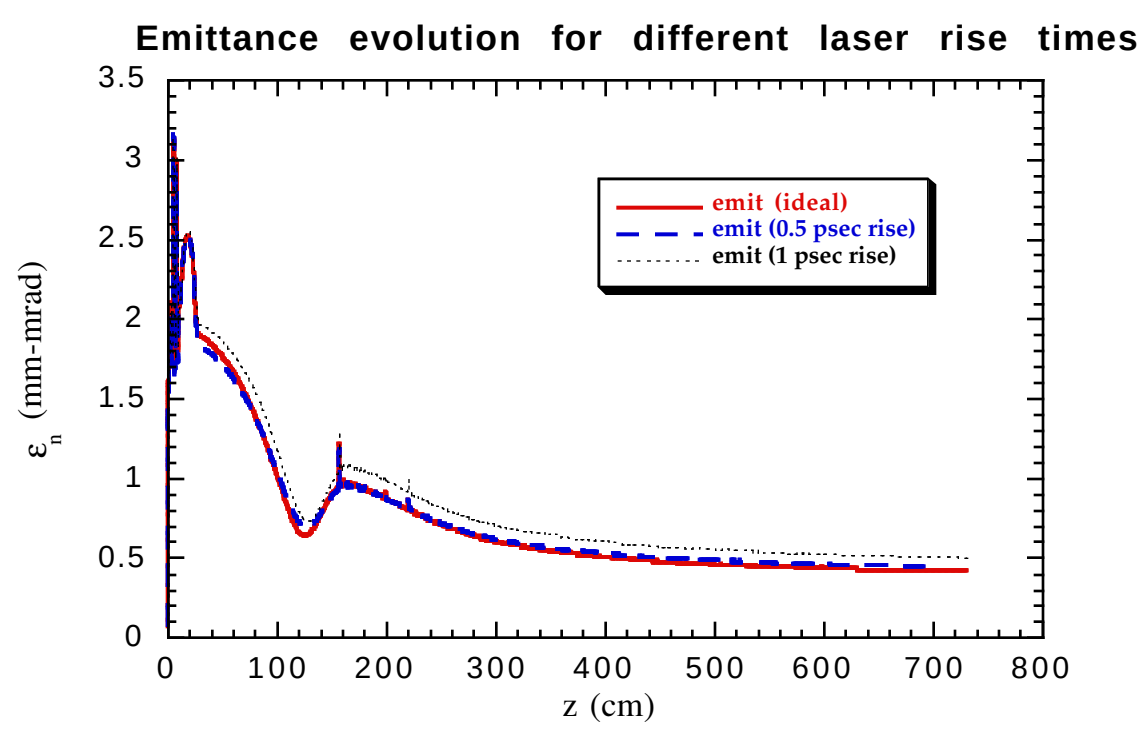

Figure 6: Emittance evolution for different laser rise time.

In the previous simulations the thermal emittance, $\varepsilon_n^{th}$, has been neglected, assuming a total emittance given by:

$$\varepsilon_n^{Total} = \sqrt{\left(\varepsilon_n^{th}\right)^2 + \left(\varepsilon_n^{cor}\right)^2}$$

With an estimated [10] thermal emittance of 0.3 mm-mrad, for a Cu cathode with UV excitation, i. e. of the same order of the space charge correlated contribution, beam dynamics could be affected by this additional effect. Preliminary computations show that the previous assumption sets only a lower limit to the total emittance, and an additional tuning of the beam line parameters is required to recover an optimized solution. Figure 7 shows the emittance evolution as computed by including 0.3 mm-mrad thermal emittance contribution, to be compared to Figure 5. The resulting total emittance is 0.4 mm-mrad, where the strength of the long solenoid has been reduced to 700 G.

# 5 CONCLUSION

By investigating a new design for the LCLS injector, a new working point for a split photoinjector has been found.

An experimental program is under way at the SLAC GTF to measure the double emittance feature along the beam line and to determine experimentally the proper location of the booster linac.

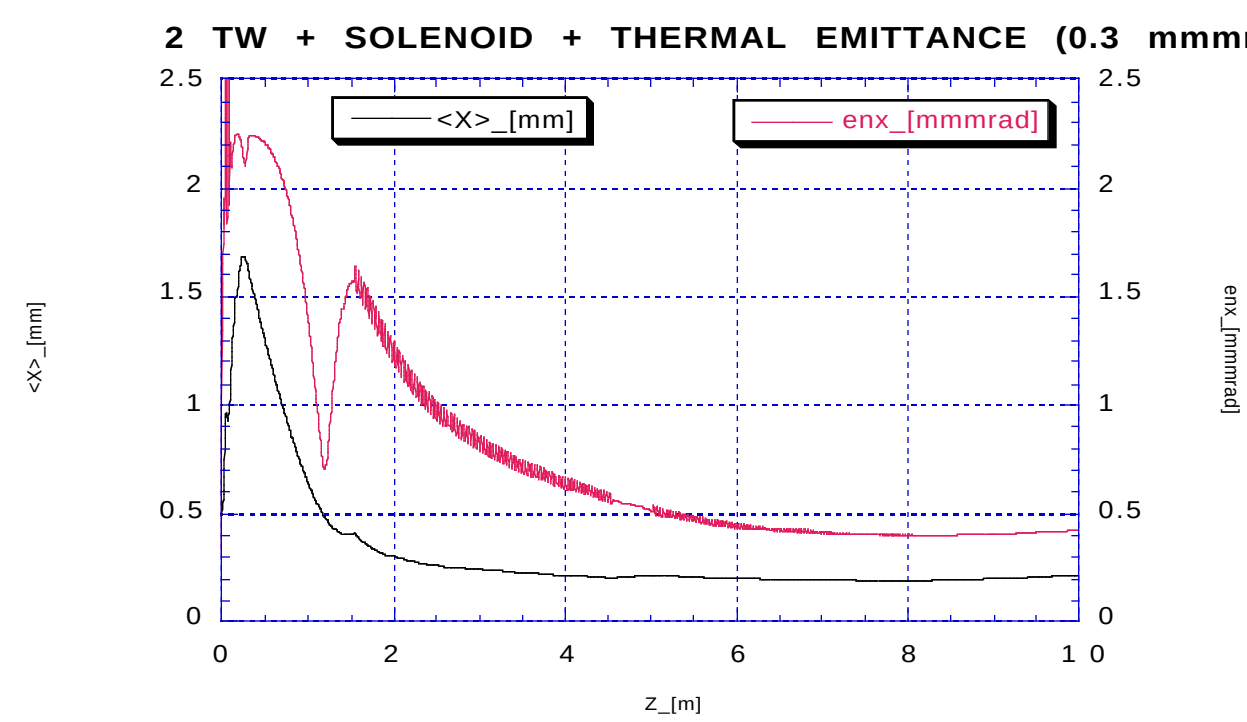

Figure 7: : Beam envelope and emittance, including 0.3 mm-mrad thermal emittance.

# 5 REFERENCES

[1] J.F. Schmerge et al., “Photocathode rf gun emittance measurements using variable length laser pulses,” SPIE 3614 (1999) 22.
[2] D.T. Palmer et al., “Emittance studies of the BNL/SLAC/UCLA 1.6 cell photocathode rf gun,” Proc. of the 1997 Particle Accelerator Conf. (1997) 2687.
[3] M. Babzien et al., "Observation of self-amplified spontaneous emission in the near-infrared and visible," Phys. Rev. E 57 (1998) 6093.
[4] R. Alley et al., “The design for the LCLS rf photo-injector”, Nucl. Instrum. and Meth. A, 429 (1999), 324.
[5] D. T. Palmer, “The next generation photoinjector”, PhD. Thesis, Stanford University (1998).
[6] M. Ferrario, A. Mosnier, L. Serafini, F. Tazzioli, J. M. Tessier, "Multi-bunch energy spread induced by beam loading in a standing wave structure", Part. Acc., **52** (1996).
[7] L. Serafini, J. B. Rosenzweig, "Envelope analysis of intense relativistic quasilaminar beams in rf photoinjectors: a theory of emittance compensation", Phys. Rev. E **55** (1997) 7565.
[8] M. Ferrario, J. E. Clendenin, D. T. Palmer, J. B. Rosenzweig, L. Serafini, “HOMDYN Study For The LCLS RF Photo-Injector”, Proc. of the 2nd ICFA Adv. Acc. Workshop on “The Physics of High Brightness Beams”, UCLA, Nov., 1999, see also SLAC-PUB-8400.
[9] UCLA version.
[10] J. Clendenin et al., “Reduction of thermal emittance of rf guns,” Proc. of the Int. Sym. on New Visions in Laser-Beam Interactions, Tokyo, Oct. 11-15, 1999 see also SLAC-PUB-8284 (1999)